\documentclass[journal]{IEEEtran}
\usepackage[utf8]{inputenc}
\usepackage{amsmath,amssymb,amsfonts}
\usepackage{algorithm} 
\usepackage{subcaption} 
\usepackage{caption}    
\usepackage{textcomp}
\usepackage{bm}
\usepackage{amsmath,amsfonts}
\usepackage{array}
\usepackage{textcomp}
\usepackage{stfloats}
\usepackage{url}
\usepackage{verbatim}
\usepackage{graphicx}
\usepackage{booktabs}

\usepackage{multirow}
\usepackage{algpseudocode}
\usepackage{diagbox}
\usepackage{balance}
\usepackage{tabularx} 
\usepackage{booktabs}   
\def\BibTeX{{\rm B\kern-.05em{\sc i\kern-.025em b}\kern-.08em
		T\kern-.1667em\lower.7ex\hbox{E}\kern-.125emX}}
\allowdisplaybreaks

\vspace{-10.pt}
\begin{document}
	\title{Efficient Beamforming Optimization for STAR-RIS-Assisted Communications: A Gradient-Based Meta Learning Approach}
	\author{Dongdong~Yang, Bin~Li,~\IEEEmembership{Member,~IEEE}, Jiguang~He,~\IEEEmembership{Senior Member,~IEEE}, Yicheng~Yan, Xiaoyu Zhang, Chongwen Huang,~\IEEEmembership{Member,~IEEE}
		\thanks{D. Yang is with Nanjing University of Information Science and Technology, Nanjing 210044, China, and also with Great Bay University, Dongguan 523000, China (e-mail: 202312200024@nuist.edu.cn).}
        \thanks{B. Li is with Nanjing University of Information Science and Technology, Nanjing 210044, China (e-mail: bin.li@nuist.edu.cn).}
        \thanks{J. He is with Great Bay University, Dongguan 523000, China (e-mail: jiguang.he@gbu.edu.cn).}
        \thanks{Y. Yan is with School of Biomedical Engineering, Guangdong Medical University, Dongguan 523808, China, and also with Great Bay University, Dongguan 523000, China (e-mail: yanyicheng@gdmu.edu.cn).}
        \thanks{X. Zhang is with the School of Electronics and Information Technology, Sun Yat-sen University, Guangzhou 510006, China, and also with Great Bay University, Dongguan 523000, China (e-mail: zhangxy988@mail2.sysu.edu.cn).}
        \thanks{Chongwen Huang is with the College of Information Science and Electronic Engineering, Zhejiang University, Hangzhou 310027, China, and also with Zhejiang Provincial Key Laboratory of Multi-Modal Communication Networks and Intelligent Information Processing, Hangzhou 310027, China (e-mail: chongwenhuang@zju.edu.cn).}
		
}
	
	\setlength{\parskip}{0pt} 
	\maketitle
	
	\vspace{-15.pt}
	\begin{abstract}
    Simultaneously transmitting and reflecting reconﬁgurable intelligent surface (STAR-RIS) has emerged as a promising technology to realize full-space coverage and boost spectral efficiency in next-generation wireless networks. Yet, the joint design of the base station precoding matrix as well as the STAR-RIS transmission and reflection coefficient matrices leads to a high-dimensional, strongly nonconvex, and NP-hard optimization problem. Conventional alternating optimization (AO) schemes typically involve repeated large-scale matrix inversion operations, resulting in high computational complexity and poor scalability, while existing deep learning approaches often rely on expensive pre-training and large network models. In this paper, we develop a gradient-based meta learning (GML) framework that directly feeds optimization gradients into lightweight neural networks, thereby removing the need for pre-training and enabling fast adaptation. Specifically, we design dedicated GML-based schemes for both independent-phase and coupled-phase STAR-RIS models, effectively handling their respective amplitude and phase constraints while achieving weighted sum-rate performance very close to that of AO-based benchmarks. Extensive simulations demonstrate that, for both phase models, the proposed methods substantially reduce computational overhead, with complexity growing nearly linearly when the number of BS antennas and STAR-RIS elements grows, and yielding up to 10 times runtime speedup over AO, which confirms the scalability and practicality of the proposed GML method for large-scale STAR-RIS-assisted communications.
	\end{abstract}
	
	\section{Introduction}
	As 5G evolves toward 6G, reconfigurable intelligent surfaces (RIS) have emerged as a key technology thanks to their programmable electromagnetics and near-passive, low-power, low-cost nature.  By adaptively adjusting the configuration of each element, RIS can reshape wireless propagation environments and enhance both spectral and energy efficiency \cite{8741198}, \cite{9424177}.  However, conventional RIS inherently operates in only one half-space, restricting signal control to one side and preventing simultaneous service to users located on opposite sides of the surface  \cite{9437234}. This fundamental limitation motivates the development of simultaneously transmitting and reflecting RIS (STAR-RIS), which can split the incident signal into controllable transmission and reflection components \cite{9570143}. With this capacity, STAR-RIS provides full-space coverage and richer beamforming degrees of freedom for next-generation wireless networks \cite{9690478}.

    Although STAR-RIS greatly expands the design flexibility of wireless systems, it also leads to a substantially more intricate optimization landscape. Compared with conventional RIS, STAR-RIS introduces additional design degrees of freedom, namely the transmission and reflection coefficients. In particular, these coefficients are intrinsically coupled through the electromagnetic structure of STAR-RIS,  including the energy conservation for the amplitude coefficients, and the fixed relative offset for the phase shift coefficients of each STAR-RIS element  \cite{9935266}.  These constraints significantly shrink the feasible region and complicate system design. Moreover, practical objectives such as maximizing the weighted sum-rate (WSR) require joint optimization of the base station (BS) precoding matrix together with the STAR-RIS amplitudes and phase shift coefficients \cite{9629335}. This leads to a highly nonconvex high-dimensional problem that is extremely challenging to solve efficiently.
    

    Numerous previous works have paid attention to these problems through various optimization methods under various scenarios. For example, Mu \emph{et al.} \cite{9570143} investigated energy-efficient beamforming for multi-user multiple-input single-output (MU-MISO) systems with minimum transmission rate constraints, and proposed a joint optimization algorithm based on semidefinite programming (SDP) and successive convex approximation, further addressing the 0-1 constraint in mode switching via a penalty method. Furthermore, Song \emph{et al.} \cite{10316600} considered multi-STAR-RIS-assisted mmWave cell-free networks, where an alternating optimization (AO) algorithm combining fractional programming, SDP, and linear quadratic relaxation was developed to jointly optimize active beamforming at mmWave base stations, passive beamforming at STAR-RIS, and user association. In the context of integrated sensing and communication (ISAC), Wang \emph{et al.} \cite{10188900} focused on scenarios with multiple eavesdroppers and proposed a low-complexity algorithm based on distance-majorization and AO methods to maximize the average received radar sensing power subject to communication quality and secrecy constraints. Moreover, considering the hardware-induced coupled-phase constraint of STAR-RIS, Wang \emph{et al.} \cite{9935266} proposed a penalty dual decomposition (PDD) framework, which achieves comparable performance to the independent-phase design with negligible performance loss.  However, these convex optimization-based methods repeatedly perform large-scale matrix inversions, leading to cubic computational complexity and infeasible updates \cite{10316600}, \cite{10475146}. Hence, lightweight and scalable beamforming designs for STAR-RIS-assisted communications are highly desirable.

    Recently, the application of deep learning (DL) in wireless communications has received considerable attention due to its inherent capability to extract valuable features from high-dimensional spaces with relatively low complexity. For example, Li \emph{et al.} \cite{10021676} investigated a joint phase shift and beamforming design for RIS-aided multiple-input multiple-output (MIMO) systems using a double DL network.  Similarly, Yuan \emph{et al.} \cite{10299716} developed a DL framework for RIS-assisted terahertz MIMO systems affected by beam squint, which utilizes mean channel covariance matrices as inputs. Beyond supervised DL, deep reinforcement learning (DRL) has also emerged as a promising tool for tackling  beamforming challenges in STAR-RIS systems. For example, Gao \emph{et al.} \cite{10901346} integrated STAR-RIS into cell-free massive MIMO systems and designed a DRL algorithm based on the soft actor-critic framework to jointly optimize the BS beamforming and STAR-RIS phase shifts, effectively accommodating user mobility. Moreover, Zhang \emph{et al.} \cite{10565781} explored STAR-RIS-assisted ISAC systems under both independent-phase and coupled-phase models, proposing a twin delayed deep deterministic policy gradient-based method. In their approach, the action space only contained reflection phase shifts, while the corresponding transmission shifts were obtained by adding or subtracting $\pi/2$ based on the agent's decision to satisfy the coupled-phase constraint.  However, a common limitation of the aforementioned data-driven DL and DRL methods is their reliance on extensive pre-training on large and often scenario-specific datasets, which incurs substantial computational cost. Furthermore, their performance is frequently confined to the particular system configurations and channel models encountered during training, lacking generalizability and scalability across diverse wireless communications.

    To address the above issues, utilizing gradients, rather than raw channel information as inputs to NNs, has emerged as a promising alternative. The gradient-driven paradigm allows networks to capture higher-order information from the optimization landscape, thereby promoting more efficient and generalizable learning. For example, Wang \emph{et al.} \cite{10620247} proposed a robust gradient-based liquid neural network framework for mmWave MIMO systems, which leverages ordinary differential equation-driven liquid neurons to design beamforming under the challenges of channel robustness and complexity. Considering RIS-assisted downlink short packet communications, Parihar \emph{et al.} \cite{11148930} developed a gradient-based DL framework to maximize the finite block length through the joint optimization of block lengths, active and passive beamforming. Furthermore, Zhu \emph{et al.} \cite{10622978} extended this concept by introducing a meta learning approach for joint active and passive beamforming in RIS-assisted mmWave MISO systems, demonstrating notable performance gains and runtime reduction compared to AO methods.  
    

Motivated by these insights, we propose a gradient-based meta learning (GML) method with extremely low computational overhead for joint beamforming optimization in STAR-RIS-assisted MU-MISO systems, applicable to both independent-phase and coupled-phase models. The main contributions are summarized as follows.
\begin{itemize}
    \item  First, we formulate the WSR maximization problem for STAR-RIS-assisted MU-MISO systems under both independent-phase and coupled-phase models, and then propose the GML method for both models. Unlike data-driven DL methods, our proposed approach directly leverages optimization gradients as NN inputs, avoiding pre-training and achieving efficient adaptation through lightweight networks.
    \item Furthermore, to address the distinct characteristics of two STAR-RIS models, we design customized loss functions. For the independent-phase scheme, the loss function is simply defined as the negative WSR. For the coupled-phase scheme, inspired by the PDD method, we incorporate a penalty term into the loss function. Specifically, the loss is defined as the sum of the negative WSR and a penalty term, where the weight of the penalty gradually increases during training. This design allows the algorithm to prioritize WSR performance in the early training stages and progressively emphasize the coupled-phase constraints later. 
    \item The proposed GML algorithm achieves performance comparable to conventional AO methods but with substantially reduced computational complexity and runtime in both STAR-RIS models. Extensive simulations demonstrate its scalability in large-scale STAR-RIS-assisted systems, where the runtime exhibits near-linear growth with system size and achieves up to 10 times speedup compared with AO.
\end{itemize}

	 \section{System Model and Problem Formulation}
In this section, the system model and the formulation of the WSR maximization problem for the STAR-RIS-assisted MU-MISO systems are presented.

As shown in Fig. 1,  we consider an MU-MISO communication system, where one BS equipped with $M$ antennas communicates with $K$ single-antenna users with the aid of an STAR-RIS comprising $N$ elements. The sets of users and STAR-RIS elements are denoted by $\mathcal{K}=\left\{ {1, \ldots ,k, \ldots ,K} \right\}$and $\mathcal{N} = \left\{ {1, \ldots ,n, \ldots ,N} \right\}$, respectively. The STAR-RIS can create full-space coverage by simultaneously transmitting and reflecting the incident signal \cite{9570143}. Since the space is divided into the transmitting area (TA) and the reflecting area (RA) by the STAR-RIS, users are divided into $\mathcal{K}_t$ and $\mathcal{K}_r$ according to their location, satisfying ${\mathcal{K}_t} \cup {\mathcal{K}_r} = \mathcal{K}$ and   ${\mathcal{K}_t}  \cap  {\mathcal{K}_r} = \emptyset $. In this paper, we assume that the direct links between the BS and the users are blocked by obstacles. Therefore, users can only communicate with the BS via the virtual links provided by the STAR-RIS. Furthermore, we assume that all the channels are known, including the channel $\mathbf{G} \in {\mathbb{C}^{N \times M}}$ between the BS and the STAR-RIS, the channel $\mathbf{H} \in {\mathbb{C}^{K \times N}}$ between the STAR-RIS and users. Therefore, the received signal at user $k \in \mathcal{K}_{\tau }, \tau  \in \{ t,r\}  $ is given by
\begin{equation}
    y_k  = \mathbf{h}_k^H{\mathbf{\Theta}  _\tau }\mathbf{G}{\mathbf{w}_k}{s_k} + \sum\limits_{j \ne k}^K {\mathbf{h}_k^H{\mathbf{\Theta} _\tau }\mathbf{G}{\mathbf{w}_j}{s_j}}  + {n_k}.
\end{equation}
Here, the transmission and reflection coefficient matrices of the STAR-RIS are given by ${\mathbf{\Theta} _\tau} = {\rm{diag}}\left( {\beta _{\tau,1}{e^{j\theta _{\tau,1}}}, \ldots ,\beta _{\tau,n}{e^{j\theta _{\tau,n}}}, \ldots ,\beta _{\tau,N}{e^{j\theta _{\tau,N}}}} \right), \tau  \in \{ t,r\}$,  where the amplitude coefficients satisfy  $\beta _{t,n},\beta _{r,n} \in [0,1]$ and $\beta _{t,n}^2 + \beta _{r,n}^2 = 1, \forall n \in \mathcal{N}$, and phase shift coefficients satisfy ${\theta _{t,n}},{\theta _{r,n}} \in [0,2\pi ),\forall n \in \mathcal{N}$. The $k$-th column of the precoding matrix $\mathbf{W} \in \mathbb{C}^{M \times K}$ at the BS is defined as $\mathbf{w}_k$, and $\mathbf{h}_k$ denotes the transpose of the $k$-th row of $\mathbf{H}$. The noise $n_k$ is the additive white Gaussian noise at user $k$ with zero mean and variance $\sigma ^2$. 

\begin{figure}[t]
    \centering
    \includegraphics[width=1\linewidth]{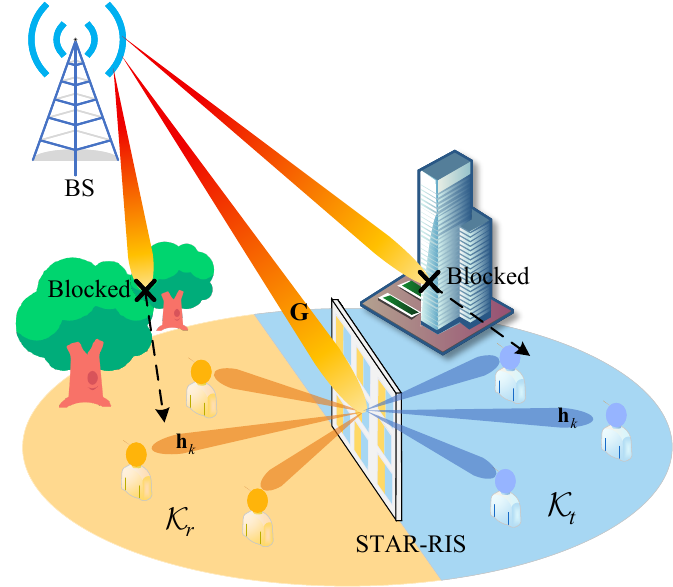}
    \caption{STAR-RIS-aided MU-MISO system.}
    \label{fig:enter-label}
\end{figure}

Notably, the total transmit power of all antennas at the BS is limited. Therefore, a constraint on $\mathbf{W}$ is introduced, given by
\begin{equation}
    {\rm Tr}\left( {{\mathbf{W}^H}\mathbf{W}} \right) \le {P_{\max }},
\end{equation}
	where $P_{\rm max}$ denotes the total maximum transmit power of the BS. Therefore, the signal-to-interference-plus-noise ratio (SINR) at user $k \in \mathcal{K}_{\tau}, \tau \in  \{ t,r\}$ is given by
\begin{equation}
    {\gamma _k} = \frac{{{{\left| {\mathbf{h}_k^{H}{\mathbf{\Theta} _\tau }\mathbf{G}{\mathbf{w}_k}} \right|}^2}}}{{\sum\nolimits_{j \ne k}^K {{{\left| {\mathbf{h}_k^H{\mathbf{\Theta} _\tau }\mathbf{G}{\mathbf{w}_j}} \right|}^2}}  + {\sigma ^2}}}.
\end{equation}

    To evaluate the system performance, the WSR serves as a metric, given by
    \begin{equation}
        R\left( {\mathbf{W},\mathbf{\Theta}_t, \mathbf{\Theta}_r } \right) = \sum\limits_{k = 1}^K {{\omega _k}{{\log }_2}\left( {1 + {\gamma _k}} \right)},
    \end{equation}
	where $\omega _k$ denotes the weight of user $k$. Therefore, the WSR maximization problem for STAR-RIS-assisted MU-MISO systems, with the joint optimization of the BS precoding matrix $\mathbf{W}$, STAR-RIS transmission, and reflection coefficient matrices $\mathbf{\Theta}_t, \mathbf{\Theta}_r$, can be formulated as
    	\begin{subequations}
		\begin{align}
			&\quad\mathop {\max }\limits_{{\bf W},{{\bm \Theta} _t},{{\bm \Theta}_r}} \;R\left( {\mathbf{W},\mathbf{\Theta}_t, \mathbf{\Theta}_r } \right)\\
			&\;{\rm{s}}{\rm{.t}}{\rm{.}}\;{\rm{tr}}\left( {{{\bf W}^H}{\bf W}} \right) \le {P_{\max }},\\
			&\;\quad\;\;\,\beta _{t,n}^2 + \beta _{r,n}^2 = 1,\forall n \in {\cal N},\\
            &\;\quad\;\;\,{\theta _{t,n}},{\theta _{r,n}} \in [0,2\pi ),\forall n \in {\cal N},\\
            &\;\quad\;\;\,\cos \left( {{\theta _{t,n}} - {\theta _{r,n}}} \right) = 0,\forall n \in {\cal N},
		\end{align}
	\end{subequations}
    where constraint (5e) denotes the coupled-phase constraint \cite{9837935}. If constraint (5e) is removed, the problem would reduce to the independent-phase model, where the transmission and reflection phase shifts of each STAR-RIS element can be independently designed.

    \section{GML Framework}
In this section, the gradient-based optimization method, the meta learning architecture, and the details of the NNs are introduced.

\subsection{Problem Reformulation} Considering that STAR-RIS introduces two coupled coefficients, namely the amplitudes and phase shifts, it is challenging to optimize them directly like conventional RIS. To address this challenge, we decompose the STAR-RIS coefficient matrices ${\bf\Theta}_t$ and ${\bf\Theta}_r$  as
\begin{equation}
    {\bf \Theta}_\tau = {\bf A}_\tau {\bf \Phi}_\tau,
\end{equation}
where ${\bf A}_{\tau} = {\rm diag}(\beta_{\tau,1}, \dots, \beta_{\tau,N}) \in {\mathbb{R}^{N \times N}}$,
${\bf \Phi}_{\tau} = {\rm diag}(e^{j\theta_{\tau,1}}, \dots, e^{j\theta_{\tau,N}}) \in {\mathbb{C}}^{N \times N}$.

However, this decomposition results in four matrices  $\{ {{\bf{A}}_t},{{\bf{A}}_r},{{\bf\Phi} _t},{{\bf\Phi} _r}\} $. Directly handling these variables would not only double the number of parameters to be optimized but also introduce significant coordination overhead when these variables are handled alternatively. More critically for our meta learning framework, it would necessitate the design of four separate sub-networks, which dramatically increase the model complexity and undermine the goal of a lightweight and efficient solution. Therefore, we aggregate transmission and reflection amplitudes as well as phase shifts via block-diagonal concatenation, given by
\begin{equation}
    {\bf A} = {\bf A}_t \oplus {\bf A}_r, \quad {\bf \Phi} = {\bf \Phi}_t \oplus {\bf \Phi}_r,
\end{equation}
where $\oplus$ denotes the block-diagonal concatenation operation. Since the coefficients for reflection users and transmission users are not shared, we introduce selection matrices, given by
\begin{equation}
    {{\bf{S}}_t} = {\rm diag}({\bf 1}_{N}, {\bf 0}_{N}), \; {{\bf{S}}_r} = {\rm diag}({\bf 0}_{N}, {\bf 1}_{N}),
\end{equation}
where ${\bf 1}_N$ and ${\bf 0}_N$ denote $N$-element all-one and all-zero vectors, \({\bf S}_t\) and \({\bf S}_r\) denote the selection matrices for transmission and reflection users, respectively. Considering that introducing the selection matrices changes the matrix dimensions, we define the following augmented channel matrices, given by
\begin{equation}
    \tilde {\bf h}_k^H = \left[ {\bf h}_k^H,{\bf h}_k^H \right]\in {\mathbb{C}^{2N\times 1}}, \quad
\tilde {\bf G} = \begin{bmatrix} {\bf G} \\ {\bf G} \end{bmatrix}\in {\mathbb{C}}^{2N \times M}.
\end{equation}
Therefore, the SINR at user $k$ can be reformulated as
\begin{equation}
    {\gamma _k} = \frac{{{{\left| {{\tilde{\bf{ h}}}_k^H{{\bf S}_\tau }{\bf A}{\bf\Phi} {\tilde{\bf{ G}}}{{\bf{w}}_k}} \right|}^2}}}{{\sum\nolimits_{j \ne k}^K {{{\left| {{\tilde{\bf{ h}}}_k^H{{\bf S}_\tau }{\bf A}{\bf \Phi} {\tilde{\bf{ G}}}{{\bf{w}}_j}} \right|}^2}}  + {\sigma ^2}}}.
\end{equation}

In summary, the original optimization problem for the transmission and reflection coefficient matrices ${\bf\Theta}_t$ and ${\bf \Theta}_r$ can be transformed into an equivalent problem with respect to the amplitude coefficient matrix ${\bf A}$ and the phase shift coefficient matrix ${\bf \Phi}$.

\subsection{Gradient-Based Meta Learning}
\subsubsection{Gradient As Input} 
Traditional data-driven DL methods typically learn an end-to-end mapping from the channel matrices $\bf H$ and $\bf G$ to the optimization variables. This makes it difficult to relate the learned mapping to the underlying optimization procedure.  In contrast, our approach uses the gradients of the WSR to the optimization variables, including the BS precoding matrix $\bf W$, STAR-RIS amplitude coefficient matrix $\bf A$, and phase shift coefficient matrix $\bf\Phi$, namely  ${\nabla _{\bf{W}}}R$, ${\nabla _{{{\bf A}}}}R$, and ${\nabla _{{{\bf\Phi}}}}R$ as input to the NNs. Then the three networks ${\rm NN}_{\bf W}(\cdot)$, ${\rm NN}_{\bf A}(\cdot)$, and ${\rm NN}_{\bf\Phi }(\cdot)$ output the updates $\Delta {\bf{W}} = {\rm NN}_{\bf W}({\nabla _{\bf W}}R)$, $\Delta {\bf A} = {\rm NN}_{\bf A}({\nabla _{\bf A}}R)$, and $\Delta {\bf \Phi}  = {\rm NN}_{\bf\Phi }({\nabla _{\bf\Phi} }R)$, which are used to refine the current matrices. This strategy inherently provides greater interpretability, as the network's behavior is explicitly tied to the optimization objective. Furthermore, by leveraging gradients, the network directly accesses first-order information from the optimization landscape, which facilitates more effective and guided parameter updates compared to learning from raw channel data alone. Thanks to the automatic differentiation mechanism of PyTorch, the gradients of each matrix to the objective function can be efficiently obtained, making the approach straightforward to implement. 


\subsubsection{Meta Learning Architecture}

Traditional data-driven meta learning methods, which require large-scale pre-training and frequent adaptation, are often unsuitable for latency-critical and dynamic scenarios. To solve this challenge, we propose a pre-training free, model-driven meta learning framework. This framework features a three-layer nested cyclic optimization structure, comprising epoch iterations, outer iterations, and inner iterations.

\paragraph{\textbf{Inner iteration}} The inner iteration is responsible for optimizing the target matrices cyclically. As shown in Fig. 2, three sub-networks are employed, namely the precoding network (PN), the amplitude network (AN), and the theta network (TN).

Within each iteration, these sub-networks update the precoding matrix $\bf W$, amplitude coefficient matrix $\bf A$, and phase shift coefficient matrix $\bf \Phi$, sequentially. In particular, the target matrix for the current sub-network is initialized from the initial, while other matrices take their most recently updated values from the preceding sub-networks. The update process in the $j$-th outer iteration can be formulated as 
\begin{equation}
    {\bf W}^{*}={\rm PN}({\bf W}^{(0,j)},{\bf A}^*,{\bf \Phi}^*),
\end{equation}
\begin{equation}
    {\bf A}^{*}={\rm AN}({\bf W}^*,{\bf A}^{(0,j)},{\bf \Phi}^*),
\end{equation}
\begin{equation}
    {\bf \Phi}^{*}={\rm TN}({\bf W}^*,{\bf A}^*,{\bf \Phi}^{(0,j)}),
\end{equation}
where ${\bf W}^{(i,j)}$, ${\bf A}^{(i,j)}$, and ${\bf \Phi}^{(i,j)}$ denote $\bf W$, $\bf A$, and $\bf\Phi$ in the $i$-th inner iteration of the $j$-th outer iteration.

\paragraph{\textbf{Outer iteration}} The outer iteration is responsible for accumulating the loss, with each iteration comprising $N_i$ inner iterations. The specific form of the loss function depends on whether the STAR-RIS operates under the independent-phase or coupled-phase model.

\noindent \textbf{• Independent-phase scheme:} For the independent-phase model, to work in a self-supervised learning manner, the loss function in $j$-th outer iteration for all sub-networks can be expressed as the negative value of the WSR, given by
\begin{equation}
    {\cal L}^j_{p, \rm ind} = -R\left( {{\bf W},{\bf A},{\bf \Phi} } \right),p \in \{ {\bf W},{\bf A},{\bf \Phi} \}.
\end{equation}
\noindent \textbf{• Coupled-phase scheme:}
For the coupled-phase model, inspired by PDD method, we reformulate the problem by introducing auxiliary variables $\tilde{{\boldsymbol\theta}}_{t}$ and $\tilde{{\boldsymbol\theta}}_{r}$ to handle the coupled-phase constraint in (5e), where the constraints on ${\bf\Theta}_{t}$ and ${\bf\Theta}_{r}$ are transferred to these auxiliary variables. Accordingly, the original problem (5) is thus equivalent to
\begin{subequations}
    \begin{align}
    &\mathop {\max }\limits_{{\bf W},{{\bm \Theta} _t},{{\bm \Theta}_r},\tilde{\boldsymbol{\theta}}_{t},\tilde{\boldsymbol{\theta}}_{r}} \;R\\
    &\;{\rm{s}}{\rm{.t}}{\rm{.}}\;\tilde \theta _{t,n}=\theta_{t,n},\tilde \theta _{r,n}=\theta_{r,n},\forall n \in {\cal N},\\
   &\;\quad\;\; \cos ({\tilde \theta _{t,n}} - {\tilde \theta _{r,n}}) = 0,\;\forall n \in {\cal N},\\
    &\;\quad\;\;\,\text{(5b)-}\text{(5d)}.
    \end{align}
\end{subequations}

To tackle the equality constraint (15b), we follow the idea of the PDD algorithm by incorporating it into the objective function (15a) as a penalty term. Thus, the problem is transformed into
\begin{subequations}
    \begin{align}
    &\mathop {\max }\limits_{{\bf W},{{\bm \Theta} _t},{{\bm \Theta}_r},\tilde{{\boldsymbol{\theta}}}_{t},\tilde{{\boldsymbol{\theta}}}_{r}} \;R-\rho\sum\nolimits_{\tau  \in \{ t,r\} } {\left\| {{{\tilde {\boldsymbol{\theta}} }_\tau } - {{\boldsymbol{\theta}} _\tau }} \right\|}^2\\
   &\;{\rm{s}}{\rm{.t}}{\rm{.}}\; \text{(5b)-}\text{(5d)}, \text{15(c)}.
    \end{align}
\end{subequations}
where $\rho$ denotes the penalty factor penalizing the violation of constraint (15b). As can be observed, when $\rho  \to \infty$, the penalty term is forced to zero, then constraint (15b) is enforced.

Since auxiliary variables $\{ {\tilde {\boldsymbol{\theta}}_t},{\tilde {\boldsymbol{\theta}}_r}\} $ only appear in the penalty term of the objective function (16a) and constraint (15c), the subproblem with respect to $\{ {\tilde {\boldsymbol{\theta}}_t},{\tilde {\boldsymbol{\theta}}_r}\} $ is given by
\begin{subequations}
    \begin{align}
    &\quad\;\mathop {\min }\limits_{\tilde{{\boldsymbol{\theta}}}_{t},\tilde{{\boldsymbol{\theta}}}_{r}} \;\rho\sum\nolimits_{\tau  \in \{ t,r\} } {\left\| {{{\tilde {\boldsymbol{\theta}} }_\tau } - {\boldsymbol{\theta} _\tau }} \right\|}^2\\
   &{\rm{s}}{\rm{.t}}{\rm{.}}\; \text{15(c)}.
    \end{align}
\end{subequations}
It can be observed that in problem (17) the auxiliary variables associated with different STAR-RIS elements are mutually independent. Thus, problem (17) can be decomposed into $N$ independent subproblems, given by
\begin{subequations}
    \begin{align}
    &\quad\;\;\mathop {\min }\limits_{{{\tilde \theta }_{t,n}},{{\tilde \theta }_{r,n}}} {\left( {{{\tilde \theta }_{t,n}} - {\theta _{t,n}}} \right)^2} + {\left( {{{\tilde \theta }_{r,n}} - {\theta _{r,n}}} \right)^2}\\
   &\quad{\rm{s}}{\rm{.t}}{\rm{.}}\; {{\tilde \theta }_{t,n}} - {{\tilde \theta }_{r,n}} =  \pm \frac{\pi }{2}\;{\rm or}\; \pm \frac{{3\pi }}{2}.
    \end{align}
\end{subequations}

Since the feasible set of problem (18) is characterized by four phase shift differences, let $t \in \left\{\pm \frac{\pi}{2},\, \pm \frac{3\pi}{2}\right\}$ denote one of the feasible values. Then, $\tilde{\theta}_{t,n}$ can be substituted by $\tilde{\theta}_{r,n} + t$, and problem (18) can be equivalently rewritten as the following  quadratic program
\begin{equation}
    \mathop {\min }\limits_{{{\tilde \theta }_{r,n}},t} \;2\tilde \theta _{r,n}^2 + 2(t-{\theta _{t,n}} - {\theta _{r,n}}){{\tilde \theta }_{r,n}}+{\left( {t - {\theta _{t,n}}} \right)^2} + \theta _{r,n}^2.
\end{equation}
By taking the first-order derivative with respect to $\tilde{\theta}_{r,n}$, the optimal solution for $\tilde{\theta}_{r,n}$ under a fixed $t$ is obtained as
$\tilde{\theta}_{r,n}^* = \frac{\theta_{t,n} + \theta_{r,n} - t}{2}$. Substituting the four possible phase shift differences $t$ and evaluating the objective in (19), the optimal solution to problem (18) for the $n$-th STAR-RIS element is given by
\begin{equation}
    \left( {{{\tilde \theta }_{t,n}^*},{{\tilde \theta }_{r,n}^*}} \right) = \mathop {\arg \min }\limits_\chi  \left\{ {\tilde \theta _{t,n}^2 + \tilde \theta _{r,n}^2 - 2{\theta _{t,n}}{{\tilde \theta }_{t,n}} - 2{\theta _{r,n}}{{\tilde \theta }_{r,n}}} \right\},
\end{equation}
The feasible set $\chi$ for the $n$-th element is
\begin{equation}
    \chi = 
\left\{
\left(
\frac{\theta_{t,n}+\theta_{r,n}+t}{2},
\frac{\theta_{t,n}+\theta_{r,n}-t}{2}
\right)
, t \in \left\{ \pm\frac{\pi}{2}, \pm\frac{3\pi}{2} \right\}
\right\},
\end{equation}
which guarantees that the global optimum of problem (17) is included. This enumeration-and-selection strategy ensures that the auxiliary variables comply with the coupled-phase constraint while minimizing the quadratic deviation from the original phase shifts.

Similarly, the loss function for the coupled-phase scheme in the $j$-th outer iteration for the TN can be expressed as the sum of the negative value of the WSR and the penalty term, given by
\begin{equation}
    {\cal L}_{\boldsymbol{\Phi} {\rm{,coupled}}}^j =  - R\left( {{\bf{W}},{\bf{A}},{\bf{\Phi }}} \right)+\rho\left\| {\boldsymbol{\theta}  - \tilde{\boldsymbol{\theta}} } \right\|^2,
\end{equation}
where ${\boldsymbol\theta}  = {\left[ {{\theta _{t,1}}, \ldots ,{\theta _{t,N}},{\theta _{r,1}}, \ldots ,{\theta _{r,N}}} \right]^T}$ and $\tilde{\boldsymbol \theta}  = {\left[ {{{\tilde \theta }_{t,1}}, \ldots ,{{\tilde \theta }_{t,N}},{{\tilde \theta }_{r,1}}, \ldots ,{{\tilde \theta }_{r,N}}} \right]^T}$. 
Moreover, the penalty factor $\rho$ plays a critical role in balancing the WSR maximization and constraint satisfaction. To ensure stable convergence, we design the training process such that the GML algorithm initially prioritizes the improvement of the WSR. Thus, $\rho$ is initialized with a small value, preventing the phase-coupled constraints from overly restricting the search space and potentially hindering performance at the early stage. As training progresses, $\rho$ is gradually increased, shifting the focus towards strictly enforcing the phase-coupling constraints. This curriculum learning strategy effectively guides the optimization to first explore high-performance regions before refining the solution to be physically feasible, ultimately ensuring strong WSR performance while adhering to the coupled-phase model. 

Considering that the coupled-phase constraint has no direct effect on the BS precoding and amplitude coefficient matrices, the loss functions for the PN and AN remain identical to those of the independent-phase scheme, given by
\begin{equation}
    {\cal L}_{{\bf A} {\rm{,coupled}}}^j =  {\cal L}_{{\bf W} {\rm{,coupled}}}^j= - R\left( {{\bf{W}},{\bf{A}},{\bf{\Phi }}} \right),
\end{equation}

\paragraph{\textbf{Epoch iteration}}
 The epoch iteration is in charge of updating the parameters of NNs, and there are $N_o$ outer iterations in each epoch iteration. After completing $N_o$ outer iterations, the losses for each NN are summed and averaged as
\begin{equation}
    \tilde {\cal L}_{p,s} = \frac{1}{{{N_o}}}\sum\limits_{j = 1}^{{N_0}} {{{\cal L}^j_{p,s}}},p \in \{ {\bf W},{\bf A},{\bf \Phi} \},s \in \{ {\rm ind},{\rm coupled}\}.
\end{equation}
The backward propagation is conducted and the Adam optimizer is used to update the NNs embedded in all sub-networks, as depicted below
\begin{equation}
    \theta _{\bf{W}}^* = {\theta _{\bf W}} + \alpha_{\bf W} {\rm Adam}\left( {{\nabla _{{\theta _{\bf W}}}}R,\tilde {\cal L}_{{\bf W},s},{\theta _{\bf W}}} \right),
\end{equation}
\begin{equation}
    \theta _{\bf{A}}^* = {\theta _{\bf A}} + \alpha_{\bf A} {\rm Adam}\left( {{\nabla _{{\theta _{\bf A}}}}R,\tilde {\cal L}_{{\bf A},s},{\theta _{\bf A}}} \right),
\end{equation}
\begin{equation}
    \theta _{\bf{\Phi}}^* = {\theta _{\bf \Phi}} + \alpha_{\bf \Phi} {\rm Adam}\left( {{\nabla _{{\theta _{\bf \Phi}}}}R,\tilde {\cal L}_{{\bf \Phi},s},{\theta _{\bf \Phi}}} \right),
\end{equation}
where $s \in \{ {\rm ind},{\rm coupled}\}$,  $\alpha_{\bf W}$, $\alpha_{\bf A}$, and $\alpha_{\bf \Phi}$ are the learning rates of the three networks. There are $N_e$ epoch iterations in the whole optimization process, and the update intervals of (26) and (27) are set to $n_1$ and $n_2$ to balance the alternative optimization. Therefore, there is one update to the parameters of the PN, AN, and TN in one, $n_1$, and $n_2$ epoch iteration, respectively. It globally controls the optimization direction of desired matrices, thus being less greedy and more efficient than the traditional AO method.

	\begin{figure}
	    \centering
	    \includegraphics[width=1\linewidth]{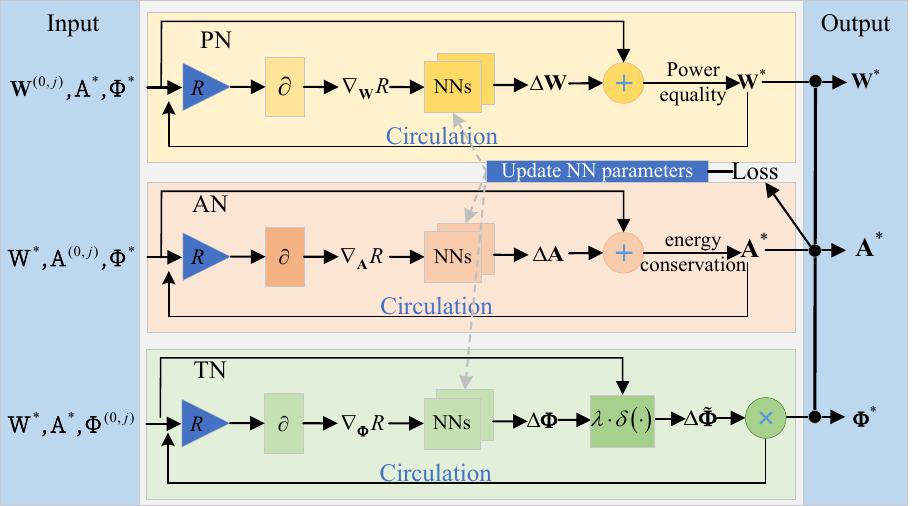}
	    \caption{The proposed GML architecture, consisting of three lightweight sub-networks PN, AN, and TN, which update $\bf W$, $\bf A$, and $\bf \Phi$ using their respective WSR gradients sequentially.}
	    \label{fig:enter-label}
	\end{figure}

\subsubsection{Precoding Network}
Given the channel information, the relation between BS precoding matrix ${\bf W}$ and the WSR is given by
\begin{equation}
    R_{\bf{W}}^{(i,j)} = \sum\limits_{k = 1}^K {{\omega _k}{{\log }_2}\left( {1 + \frac{{{{\left| {{\bf{h}}_k^H{{\hat {\bf{\Theta }}}_\tau }{\bf{G}}{\bf w}_k^{(i,j)}} \right|}^2}}}{{\sum\nolimits_{l \ne k}^K {{{\left| {{\bf{h}}_k^H{{\hat {\bf{\Theta }}}_\tau }{\bf{G}}{\bf w}_l^{(i,j)}} \right|}^2} + {\sigma ^2}} }}} \right)},
\end{equation}
where $\hat{\bf \Theta}_{\tau}$ is either initialized or updated STAR-RIS coefficient matrix, which can be calculated based on (6). In the $j$-th outer and $i$-th inner iteration, the WSR $R_{\bf W}^{(i,j)}$ is computed first, and the gradient of ${\bf W}^{(i,j)}$ with respect to the WSR ${\nabla _{\bf{W}}}R$ is then fed into the NNs. The network outputs an update $\Delta {\bf W}^{(i,j)}$, which is added to ${\bf W}^{(i,j)}$. However, the updated precoding matrix $\bf W$ may violate the transmit power constraint (2). As demonstrated in \cite{10623434}, any nontrivial stationary point $\bf W$ must conform to the power constraint with equality. Therefore, the result can be directly regulated to satisfy the power equality constraint, given by
\begin{equation}
    {\bf W}^* = {{\bf W}^{(i,j)}} + \Delta {{\bf W}^{(i,j)}},
\end{equation}
\begin{equation}
    {{\bf W}^{(i + 1,j)}} = \sqrt {\frac{{{P_{\max }}}}{{{\rm Tr}\left( ({\bf W}^*)^H {\bf W}^* \right)}}} {{\bf W}^*}.
\end{equation}

\subsubsection{Amplitude Network}
Similar to (28), the WSR in the $i$-th inner iteration and $j$-th outer iteration is denoted as $R_{\bf A}^{(i,j)}$, which can be expressed as
\begin{equation}
    R_{\bf A}^{(i,j)} = \sum\limits_{k = 1}^K {{\omega _k}{{\log }_2}\left( {1 + \frac{{{{\left| {\tilde{\bf h}_k^H{{\bf S}_\tau }{{\bf A}^{(i,j)}}\hat{\bf \Phi} \tilde{\bf G}{\hat{\bf w}_k}} \right|}^2}}}{{\sum\nolimits_{l \ne k}^K {{{\left| {\tilde{\bf h}_k^H{{\bf S}_\tau }{{\bf A}^{(i,j)}}\hat{\bf \Phi} \tilde{\bf G}{\hat{\bf w}_l}} \right|}^2} + {\sigma ^2}} }}} \right)} ,
\end{equation}
where  $k \in {\cal K}_{\tau}, \tau \in \{ t,r\} $, ${\hat{\bf \Phi}}$ and ${\hat{\bf w}_k}$ are initialized or updated matrices. Similar to PN, the gradient of ${\bf A}^{(i,j)}$ with respect to the WSR ${\nabla _{\bf{A}}}R$ is fed to the NN, then the output $\Delta {\bf A}^{(i,j)}$ is added to  ${\bf A}^{(i,j)}$, given by
\begin{equation}
    {\bf A}^{*} = {\bf A}^{(i,j)} + \Delta {\bf A}^{(i,j)}.
\end{equation}
However, the obtained ${\bf A}^*$ may not satisfy the energy conservation constraint (5c). To address this, we first introduce a transformation matrix $\bf M$, given by
\begin{equation}
    {\bf M} = \left[ {\begin{array}{*{20}{c}}
{{{\bf 0}_{N \times N}}}&{{{\bf I}_{N \times N}}}\\
{{{\bf I}_{N \times N}}}&{{{\bf 0}_{N \times N}}}
\end{array}} \right],
\end{equation}
which is composed of four $N \times N$ matrices, where the diagonal blocks are zero matrices and the off-diagonal  blocks are identity matrices. With this transformation, the reflection and transmission amplitude coefficients in ${\bf A}^*$ can be exchanged, expressed as 
\begin{equation}
    {{\bar {\bf A}}^*} = {{\bf M}^T}{{\bf A}^*}{\bf M},
\end{equation}
Consequently, the amplitude coefficient matrix $\bf A$ satisfying Constraint (5c) can be calculated by 
\begin{equation}
    {{\bf A}^{(i + 1,j)}} = {\left( {{{({{\bf A}^*})}^T}{{\bf A}^*} + {{({{\bar {\bf A}}^*})}^T}{{\bar {\bf A}}^*}} \right)^{ - \frac{1}{2}}}{{\bf A}^*}.
\end{equation}
Although (35) involves matrix inversion operations, the actual computational complexity remains ${\cal O}(N)$ as all the matrices are diagonal.

\subsubsection{Theta Network}
The WSR for TN can be expressed as 
\begin{equation}
    R_{\bf \Phi} ^{(i,j)} = \sum\limits_{k = 1}^K {{\omega _k}{{\log }_2}\left( {1 + \frac{{{{\left| {\tilde{\bf{h}}_k^H{{\bf{S}}_\tau }{\bf{\hat A}}{{\bf\Phi} ^{(i,j)}}\tilde{\bf{G}}{{\hat {\bf{w}}}_k}} \right|}^2}}}{{\sum\nolimits_{l \ne k}^K {{{\left| {\tilde{\bf{h}}_k^H{{\bf{S}}_\tau }{\bf{\hat A}}{{\bf \Phi} ^{(i,j)}}\tilde{\bf{G}}{{\hat {\bf{w}}}_l}} \right|}^2} + {\sigma ^2}} }}} \right)},
\end{equation}
where $\hat{\bf A}$ and $\hat{\bf w}_k$ are initialized or updated matrices. Although (36) shares a similar formulation with (28) and (31), the behavior of the target optimization variable is quite different, which presents a major challenge. Specifically, if the phase shift coefficient directly adopts the update method like these in PN and AN, namely ${\theta_{\tau,n} ^{(i + 1,j)}} = {\theta_{\tau,n} ^{(i,j)}} + \Delta {\theta_{\tau,n} ^{(i,j)}}$, the change in ${\bf \Phi}^{(i+1,j)}$ with  $\Delta {\theta_{\tau,n} ^{(i,j)}}$ would not be monotonic due to the periodicity of trigonometric functions. Consequently, the output of the NN may exceed its intended range, resulting in fluctuations in the WSR and further difficulty in achieving convergence. This may pose a great challenge in determining the optimal point, which is crucial to address in the design of TN. In light of the fact that the period of a trigonometric function is $2\pi$, we design a customized regulator to ensure that $\Delta\theta_{\tau,n}$ is constrained in $[0,2\pi )$, given by
\begin{equation}
    \Delta {{\tilde \theta }_{\tau ,n}} = \lambda  \cdot \delta \left( {\Delta {\theta _{\tau ,n}}} \right),
\end{equation}
where $\lambda$ is an amplification operator and $\delta \left(  \cdot  \right)$ denotes the sigmoid function. This design ensures that the phase shift coefficient matrix $\bf \Phi$ can be constrained within a limited range in each inner iteration, which can be expressed as 
   \begin{equation}
   \begin{array}{l}
         {{\bf{\Phi }}^{(i + 1,j)}}= {\rm{diag}}\left( {{e^{j\left( {\theta _{\tau ,1}^{(i,j)} + \Delta {{\tilde \theta }_{\tau ,1}}} \right)}}, \ldots ,{e^{j\left( {\theta _{\tau ,N}^{(i,j)} + \Delta {{\tilde \theta }_{\tau ,N}}} \right)}}} \right)  \\
       \quad\quad\quad\;\;\, = {{\bf{\Phi }}^{(i,j)}} \cdot \Delta \tilde {\bf{\Phi }}.
   \end{array}
   \end{equation}
   The detailed parameters for all NNs are listed in Table I, and the algorithm is summarized in Algorithm I.

\begin{algorithm}[t]
\caption{GML Workflow}
\label{alg:GMML}
\begin{algorithmic}[1]
    \State Input channel matrices $\mathbf{H}, \mathbf{G}$.
    \State Randomly Initialize $\theta_{\bf W}, \theta_{\bf A}, \theta_{\bf \Phi},\mathbf{W}^{(0,1)}, \mathbf{A}^{(0,1)}, \mathbf{\Phi}^{(0,1)}$.
    \State Normalize $\mathbf{W}^{(0,1)}$ and $\mathbf{A}^{(0,1)}$ by (30) and (35).
    \State Set maximum WSR $R_{\max}=0$.
    \For{$k \gets 1,2,\cdots,N_e$}
        \State $\tilde{\mathcal L}_{p,s} \gets 0$, $p \in \{\mathbf W,\mathbf A,\bm{\Phi}\}$, $s \in \{\mathrm{ind},\mathrm{coupled}\}$;
		
        \For{$j \gets 1,2,\cdots,N_o$}
            \State $\mathbf{W}^{(0,j)} = \mathbf{W}^{(0,1)}$, $\mathbf{A}^{(0,j)} = \mathbf{A}^{(0,1)}$,$\mathbf{\Phi}^{(0,j)} = \mathbf{\Phi}^{(0,1)}$.
            \For{$i \gets 1,2,\cdots,N_i$}
                \State $R_{\bf W}^{(i-1,j)} = R(\mathbf{W}^{(i-1,j)}, \mathbf{A}^{*},\mathbf{\Phi}^{*})$;
                \State $\Delta \mathbf{W}^{(i-1,j)} = \text{NN}_{\bf W}(\nabla_{\bf W} R_{\bf W}^{(i-1,j)})$;
                \State Update ${\bf W}^{(i,j)}$ by (29) and (30);
            \EndFor
            \State $\mathbf{W}^* = \mathbf{W}^{(N_i,j)}$;
            \For{$i \gets 1,2,\cdots,N_i$}
                \State $R_{\bf A}^{(i-1,j)}=R({\bf W}^{*},{\bf A}^{(i-1,j)},{\bf \Phi}^*)$;
                \State $\Delta \mathbf{A}^{(i-1,j)} = \text{NN}_{\bf A}(\nabla_{\bf A} R_{\bf A}^{(i-1,j)})$;
                \State Update ${\bf A}^{(i,j)}$ by (32) and (35).
            \EndFor
                \State ${\bf A}^*={\bf A}^{(N_i,j)}$;
            \For{$i \gets 1,2,\cdots,N_i$}
                \State $R_{\bf \Phi}^{(i-1,j)}=R({\bf W}^{*},{\bf A}^*,{\bf \Phi}^{(i-1,j)})$;
                \State $\Delta \mathbf{\Phi}^{(i-1,j)} = \text{NN}_{\bf \Phi}(\nabla_{\bf \Phi} R_{\bf \Phi}^{(i-1,j)})$;
                \State Update $\mathbf{\Phi}^{(i,j)}$ by (37) and (38);
            \EndFor
             \State $\mathbf{\Phi}^* = \mathbf{\Phi}^{(N_i,j)}$;
            \State Calculate loss $\mathcal{L}^j_{p,s}$ by (14) and (22) with $\mathbf{W}^*$, $\mathbf{A}^*$,  and $\mathbf{\Phi}^*$;
            \State $\tilde{\mathcal{L}}_{p,s} = \tilde{\mathcal{L}}_{p,s} + \mathcal{L}^j_{p,s}$.
            \If {$R(\mathbf{W}^*, \mathbf{A}^*,{\bf \Phi}^*) > R_{\max}$} 
                \State $R_{\max} = R(\mathbf{W}^*, \mathbf{A}^*,{\bf \Phi}^*)$;
                \State $\mathbf{W}^{\rm opt} = \mathbf{W}^*$, $\mathbf{A}^{\rm opt} = \mathbf{A}^*$, $\mathbf{\Phi}^{\rm opt} = \mathbf{\Phi}^*$.
            \EndIf
        \EndFor
        \State $\tilde{{\mathcal{L}}}_{p,s} = \frac{1}{N_o} \tilde{\mathcal{L}}_{p,s}$;
        \State Update $\theta_{\bf W}$ as (25).
        \If{${k}\,\bmod \,{n_1} = 0$}
            \State Update $\theta_{\bf A}$ as (26).
        \EndIf
        \If{${k}\,\bmod \,{n_2} = 0$}
            \State Update $\theta_{\bf \Phi}$ as (27).
        \EndIf
    \EndFor
    \State Calculate $\mathbf{\Theta}_t^{\rm opt}$ and $\mathbf{\Theta}_r^{\rm opt}$ with ${\bf A}^{\rm opt}$ and ${\bf \Phi}^{\rm opt}$.
    \State \Return $\mathbf{W}^{\rm opt}, \mathbf{\Theta}_t^{\rm opt},\mathbf{\Theta}_r^{\rm opt}$.
\end{algorithmic}
\end{algorithm}

\newcolumntype{Y}{>{\centering\arraybackslash}X} 

\begin{table}[t]
\caption{Parameters of the GML Method}
\centering
\setlength{\tabcolsep}{4pt} 
\renewcommand{\arraystretch}{1.2} 
\begin{tabularx}{0.48\textwidth}{cXXX}
\toprule
\textbf{} & \textbf{PN} & \textbf{AN} & \textbf{TN} \\
\midrule
Input Layer            & $M$ & $2\times N$ & $2\times N$ \\
Linear Layer           & 200          & 300         & 300 \\
ReLU Layer             & 200          & 300         & 300 \\
Output Layer           & $M$ & $2\times N$ & $2\times N$ \\
Differential Regulator & /            & /           & $2\times N$ \\
\bottomrule
\end{tabularx}

\end{table}

\section{Complexity Analysis}
   In this section, the computational complexity of the proposed GML method is detailed. First, we analyze the computational complexity within an inner iteration for each sub-network. 

   The analysis begins by examining the PN, whose computational complexity mainly arises from calculating the WSR and the NNs. For calculating WSR, we mainly rely on (3) and (4). Considering that the dimension of ${\bf h}_k^H$ is $1 \times N$ and ${\bf \Theta}_{\tau}$ is a diagonal matrix with dimension $N \times N$, the computational complexity for multiplying ${\bf h}_k^H$ with ${\bf \Theta}_{\tau}$ is ${\cal O}(N)$. The matrix $\bf G$ with dimension $N \times M$ adds a complexity of  ${\cal O}(MN)$, and the vector ${\bf w}_k$ adds a complexity of ${\cal O}(M)$ when multiplied, respectively. Therefore, the complexity of computing ${\bf h}_k^H{\bf \Theta}_{\tau}{\bf G}{\bf w}_k$ is ${\cal O}(MN)$. This would be performed $K$ times to compute the SINR while ${\bf h}_k^H{\bf \Theta}_{\tau}{\bf G}$ is kept fixed, resulting in the total computational complexity being ${\cal O}(MN+KM)={\cal O}(MN)$ . The calculation of the WSR involves calculating the SINR $K$ times, leading to the computational complexity being ${\cal O}(KMN)$. Thanks to the automatic differentiation of PyTorch, the computation of the gradient for ${\bf W}$ is carried out concurrently, not contributing mainly to the total complexity. Since embedded NNs are relatively small and shallow, the computational complexity of NNs is approximately ${\cal O}(M)$. In particular, the output of the NNs, namely ${\bf W}^{*}$, need to be normalized by (30), whose complexity is ${\cal O}(K^2M)$. Therefore, the complexity in a single inner iteration within the PN is ${\cal O}(KMN+M+K^2M)={\cal O}(KMN)$.

For the AN, different from the computational complexity of the PN, the computation of the SINR is based on (10). This may cause the computation cost to be ${\cal O}(2N+4MN+KM)$, which must be computed $K$ times to calculate the WSR. Furthermore, there is also a normalization for the output of the NNs based on (35), whose computational cost is ${\cal O}(4N^2)$. Therefore, the total computational complexity of a single inner iteration within the AN is ${\cal O}(K(2N+4MN+KM)+4N^2)={\cal O}(KMN+N^2)$.

For the TN, the way of computing the WSR is  same as that of AN, and the output of it does not need to be normalized. Therefore, the computational complexity of the TN is ${\cal O}(K(2N+4MN+KM))={\cal O}(KMN)$.

Second, considering the number of inner, outer, and epoch iterations, the overall complexity of the proposed GML method is $N_{e}N_{o} N_{i}({\cal O}(KMN)+{\cal O}(KMN+N^2)+{\cal O}(KMN))={\cal O}(N_{e}N_{o} N_{i}(KMN+N^2))$.  Correspondingly, the overall computational complexity of the AO method per iteration is 
${\cal O}\!\left(L_3 \left(L_1 K^3 M^3 + L_2 N^3\right)\right)$, 
where $L_1$, $L_2$, and $L_3$ denote the numbers of iterations \cite{9935266}. This further demonstrates that the proposed GML algorithm achieves lower complexity and higher efficiency compared to the conventional AO algorithm, particularly in scenarios with large-scale antennas and a massive number of STAR-RIS elements. 
\begin{figure}
    \centering
    \includegraphics[width=1\linewidth]{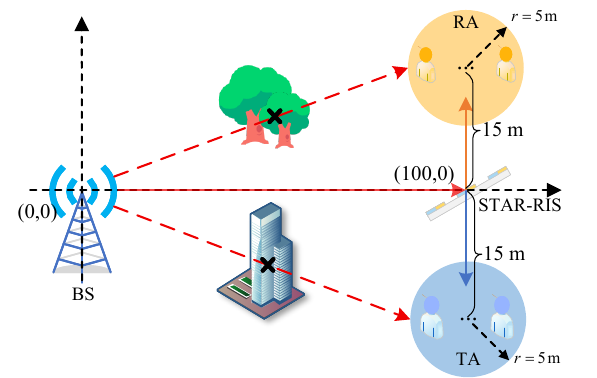}
    \caption{The simulation scenario for STAR-RIS-assisted MU-MISO communications.}
    \label{fig:placeholder}
\end{figure}

\section{simulation results}

In this section, simulation results are presented to demonstrate the performance of the proposed GML method. As shown in Fig. 3, we assume that all devices are located at the same altitude, where the BS and the STAR-RIS are located at $(0,0)$ meters and $(100,0)$ meters, respectively. The users in the RA and TA are randomly distributed in circular areas with a radius of $5$ meters centered at $(100,15)$ meters and $(100,-15)$ meters, respectively. The detailed simulation parameters are summarized in TABLE II, which are almost same as those in \cite{8982186}. In particular, the path loss is set according to the 3GPP propagation environment.

In this paper, all channels, including the channel between the BS and the STAR-RIS and the channel between the STAR-RIS and users, follow the Rician fading, modeled as
	\begin{equation}
		{\bf{G}} = {L _{\bf{G}}} \left( {\sqrt {\frac{{{K_{\bf G}}}}{{1 + {K_{\bf G}}}}} {{\bf G}^{\rm LoS}} + \sqrt {\frac{1}{{1 + {K_{\bf G}}}}} {{\bf G}^{\rm NLoS}}} \right),
	\end{equation}
\begin{equation}
		{\bf{h}}_k = {L _{\bf{h}}}\left( {\sqrt {\frac{{{K_{\bf h}}}}{{1 + {K_{\bf h}}}}} {{\bf h}_k^{\rm LoS}} + \sqrt {\frac{1}{{1 + {K_{\bf h}}}}} {{\bf h}^{\rm NLoS}}} \right),
	\end{equation}
where $L_{\bf G}$ and $L_{\bf h}$ denote the corresponding path losses, $K_{\bf G}$ and $K_{\bf h}$ are the Rician factor and we set $K_{\bf G}=K_{\bf h}=10$. ${\bf G}^{\rm LoS}$ and ${\bf h}_k^{\rm LoS}$ are the line-of-sight (LoS) components, while ${\bf G}^{\rm NLoS}$ and ${\bf h}_k^{\rm NLoS}$ are the non-LoS (NLoS) components whose elements are chosen from ${\cal CN}(0,1)$. 

\begin{table}[t]
\centering
\caption{Simulation Parameters}
\renewcommand{\arraystretch}{1.2} 
\begin{tabular}{|c|c|c|c|}
\hline
\textbf{Parameters} & \textbf{Values} & \textbf{Parameters} & \textbf{Values} \\
\hline
$N$ & 100 &  $M$ & 64 \\
\hline
$K$ & 4 & Path loss (dB) & $35.6 + 22.0 \lg d$ \\
\hline
$P_{\max}$ & 10 dBm & Noise power $\sigma^2$ & $-80$ dBm \\
\hline
\end{tabular}
\end{table}
In the simulations, we compare our results with several others on $N_s$ independent channel samples. The system parameters $N_s$, $N_e$, $N_o$, $N_i$, $\alpha_{\bf W}$, $\alpha_{{\bf\Theta}_t}$, $\alpha_{{\bf\Theta}_r}$, $\lambda$, $n_1$, and $n_2$ are set as $100$, $500$, $1$, $1$, $1\times 10^{-3}$, $5\times10^{-3}$, $5\times10^{-3}$, $2\pi$, 5 and 5, respectively. We run the simulation on a computer equipped with an Intel Core i5-8300H CPU and a GTX 1050Ti GPU using PyTorch 1.8.0 and Python  3.7.16. To demonstrate the performance of the proposed method, the following benchmark schemes are included.
	\begin{itemize}
        \item \textbf{AO scheme with independent phase:} In this scheme, the block coordinate descent  and weighted minimum mean square error  methods are utilized to optimize $\bf W$, ${\bf \Theta}_t$, and ${\bf \Theta}_r$, alternatively.
        \item \textbf{AO scheme with coupled phase:} In this scheme, the optimization method in \cite{9935266} is adopted, which utilizes the PDD algorithm to address the phase-coupled constraint.
		\item \textbf{Random phase scheme:} In this scheme, ${\bf\Theta}_t$ and ${\bf\Theta}_r$ are randomly initialized, and $\bf W$ is optimized by GML.
		\item \textbf{Conventional RIS:} In this scheme, there are two $\frac{N}{2}$-element reflect-only and transmit-only RISs deployed adjacent to each other.
	\end{itemize}

\begin{figure}
    \centering
    \includegraphics[width=1\linewidth]{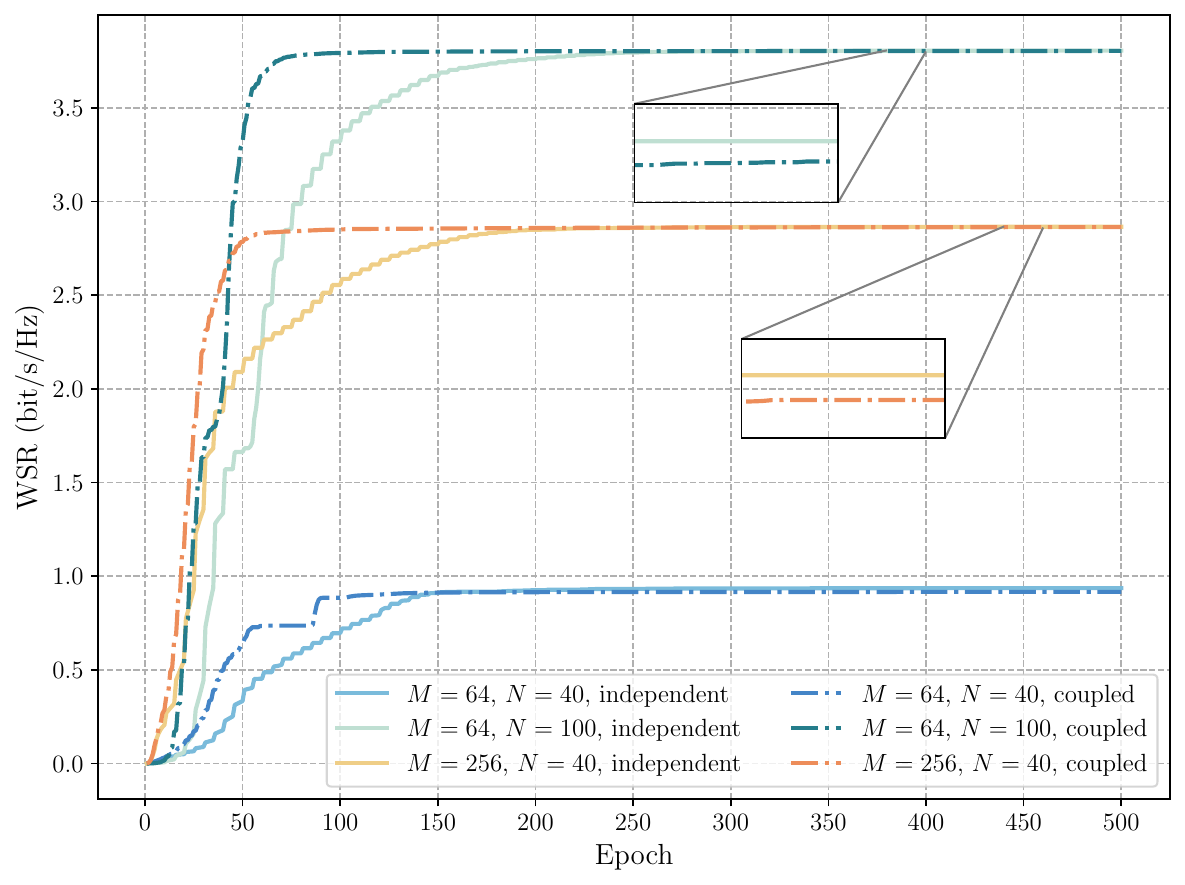}
    \caption{WSR vs. Epoch.}
    \label{fig:enter-label}
\end{figure}

In Fig. 4, we compare the convergence performance of the proposed GML algorithm under different numbers of STAR-RIS elements $N$ and BS antennas $M$. The results indicate that the proposed method consistently achieves fast convergence within approximately $150$ epochs across all configurations. Moreover, larger antenna arrays and more STAR-RIS elements yield higher WSR, confirming the scalability and efficiency of the proposed algorithm. In addition, it can be clearly observed that under the coupled-phase constraint, the proposed GML method still achieves a WSR very close to that of the independent phase shift scheme.  
\begin{figure}
    \centering
    \includegraphics[width=1\linewidth]{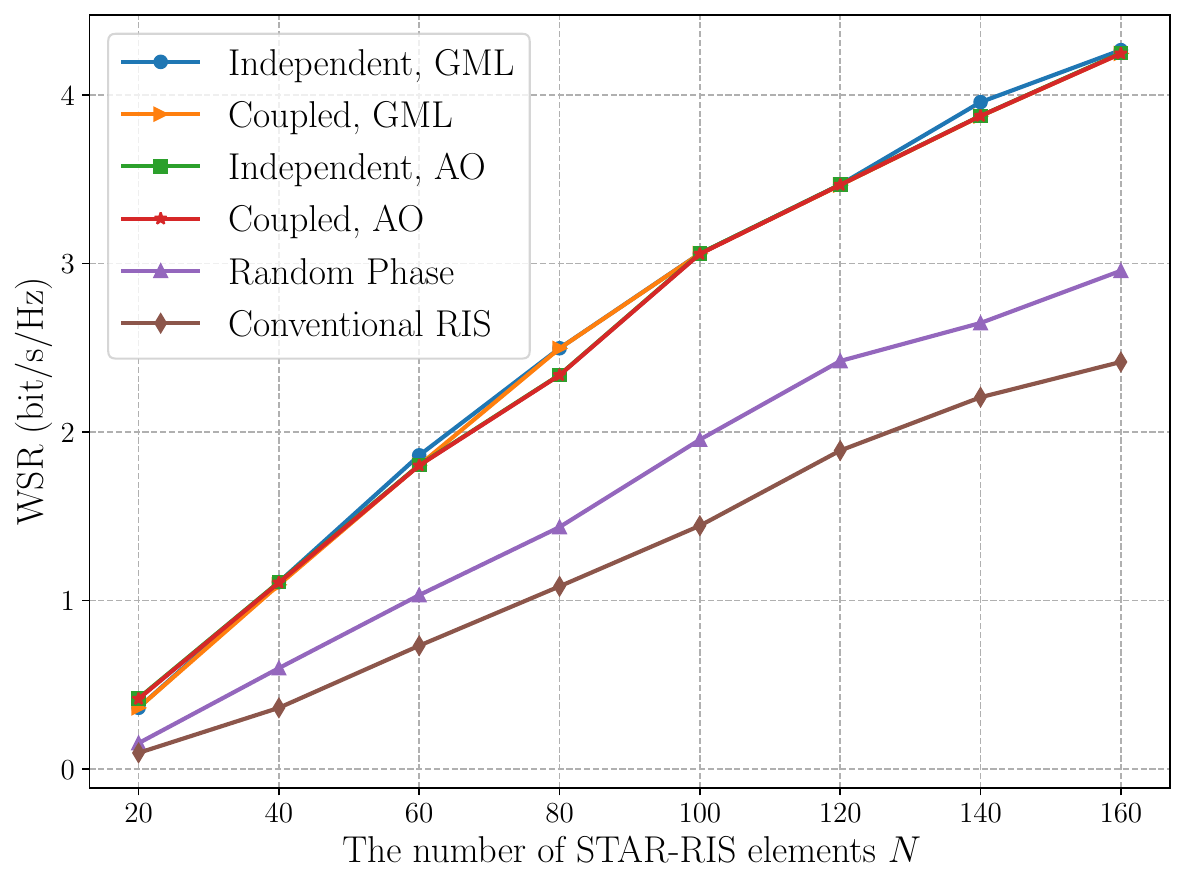}
    \caption{WSR vs. the number of STAR-RIS elements.}
    \label{fig:enter-label}
\end{figure}
\begin{figure}[t]
	\centering
	\includegraphics[width=1\linewidth]{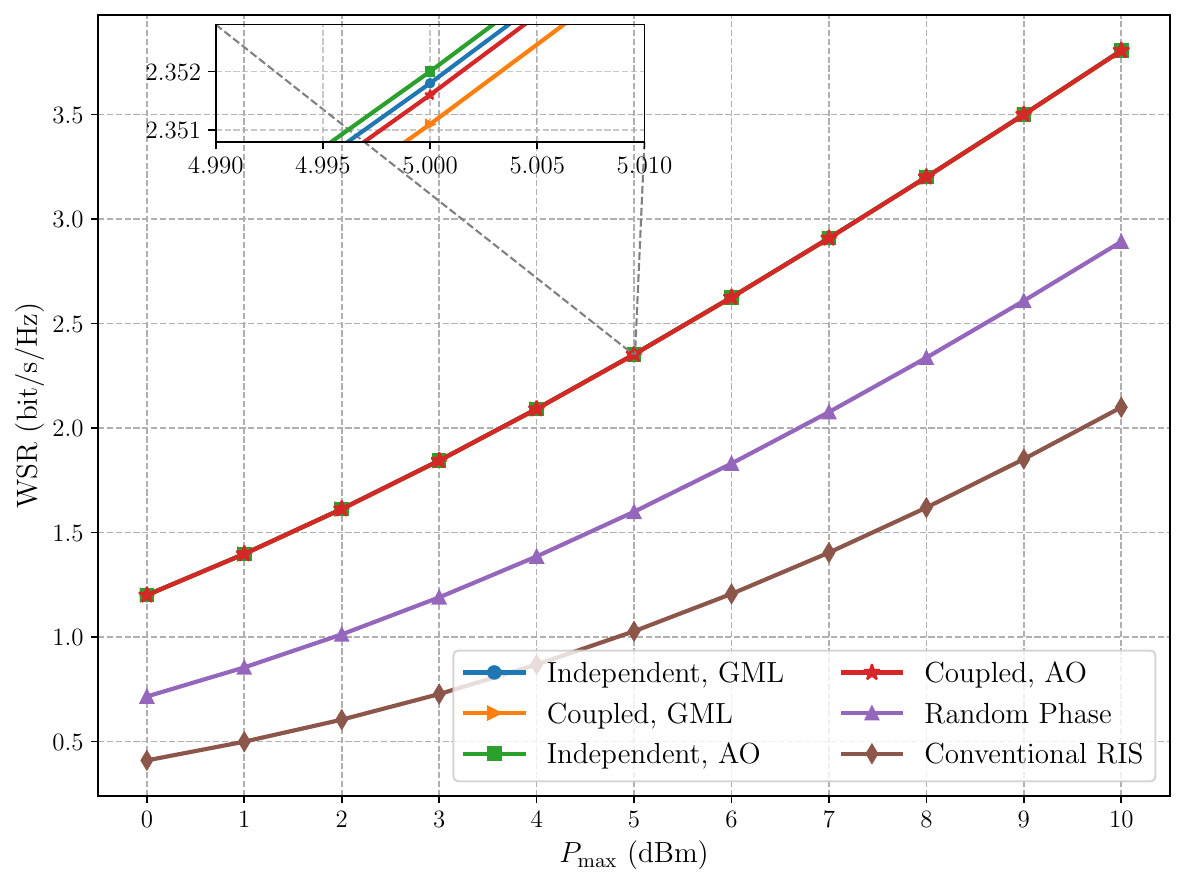}
	\caption{WSR vs. transmit power.}
	\label{fig:enter-label}
\end{figure}
\begin{figure}[t]
    \centering
    \includegraphics[width=1\linewidth]{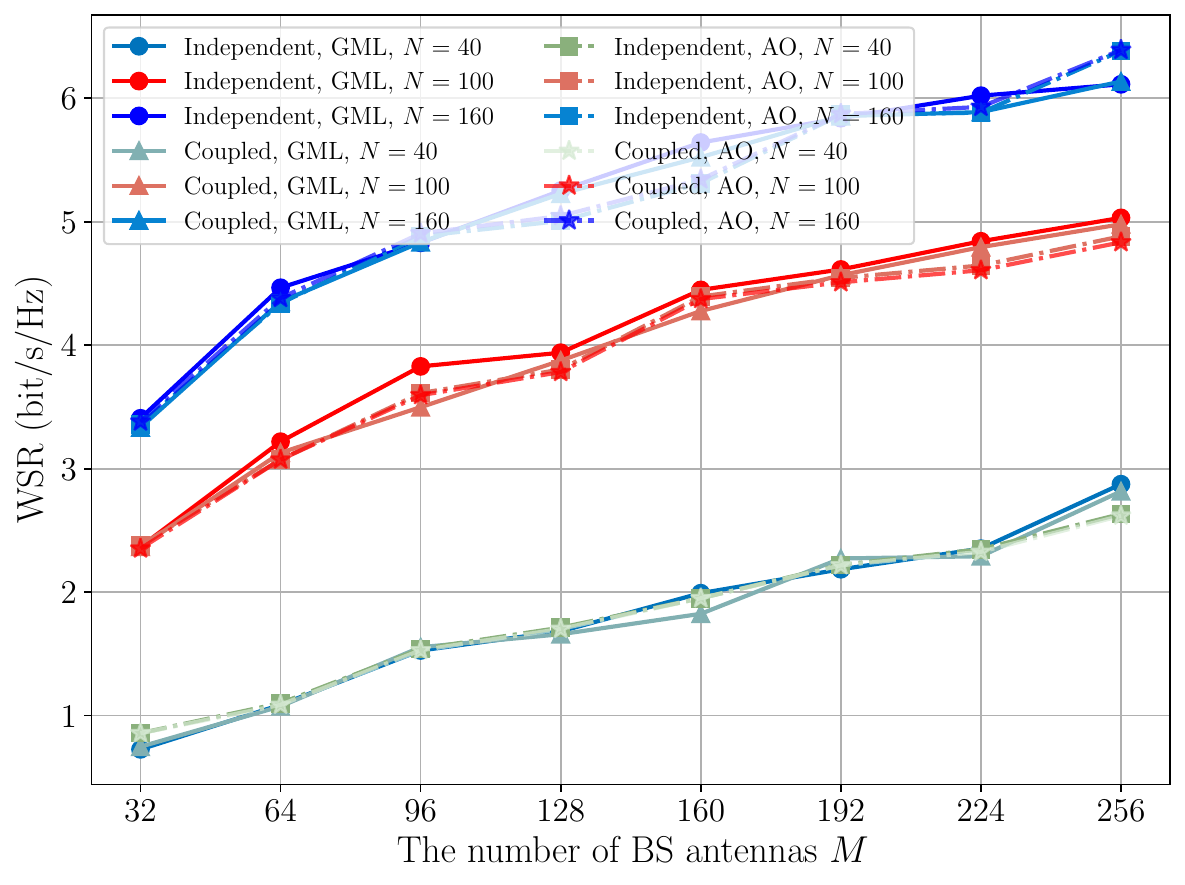}
    \caption{WSR vs. the numbers of BS antennas and STAR-RIS elements.}
    \label{fig:enter-label}
\end{figure}
\begin{figure}
    \centering
    \includegraphics[width=1\linewidth]{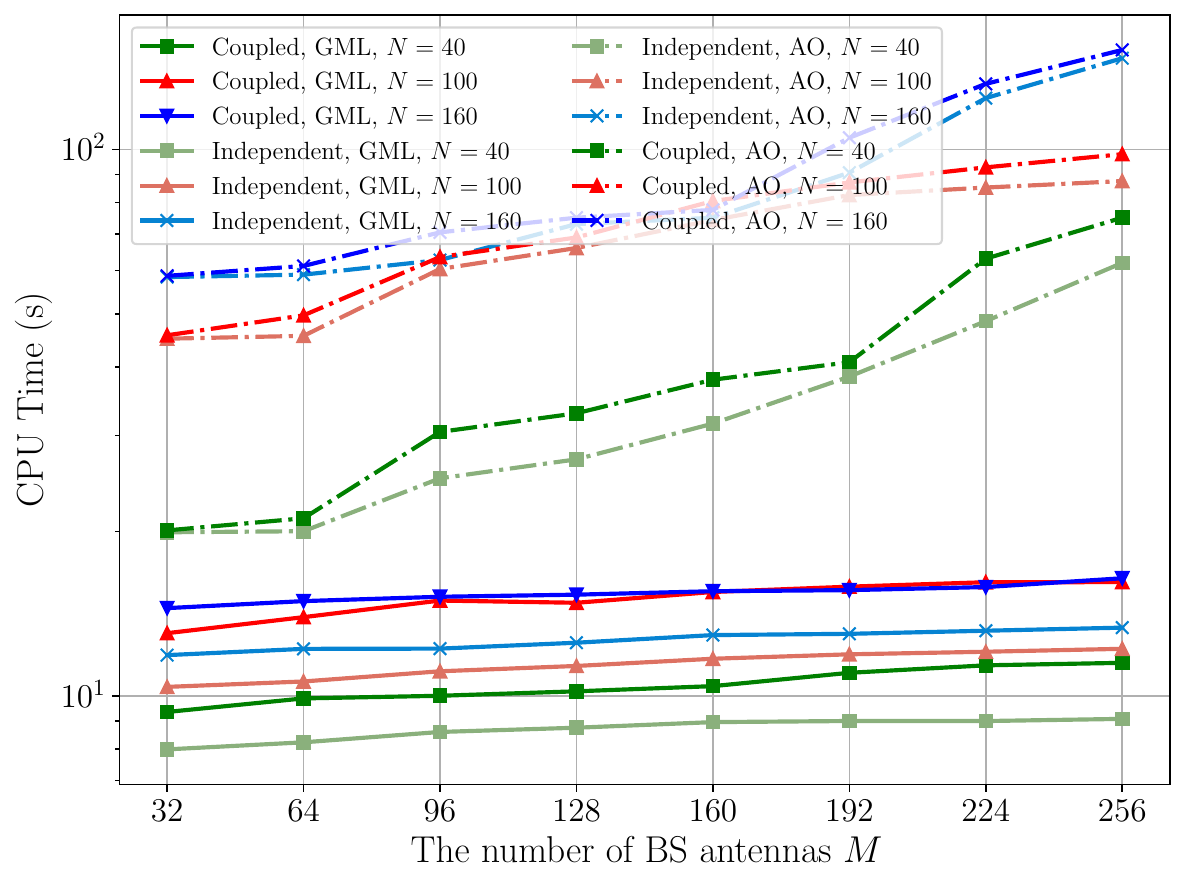}
    \caption{Average CPU time vs. the numbers of BS antennas and STAR-RIS elements.}
    \label{fig:enter-label}
\end{figure}

\begin{figure}[t]
    \centering
    \begin{subfigure}[htbp]{0.45\textwidth}
        \centering
        \includegraphics[width=\textwidth]{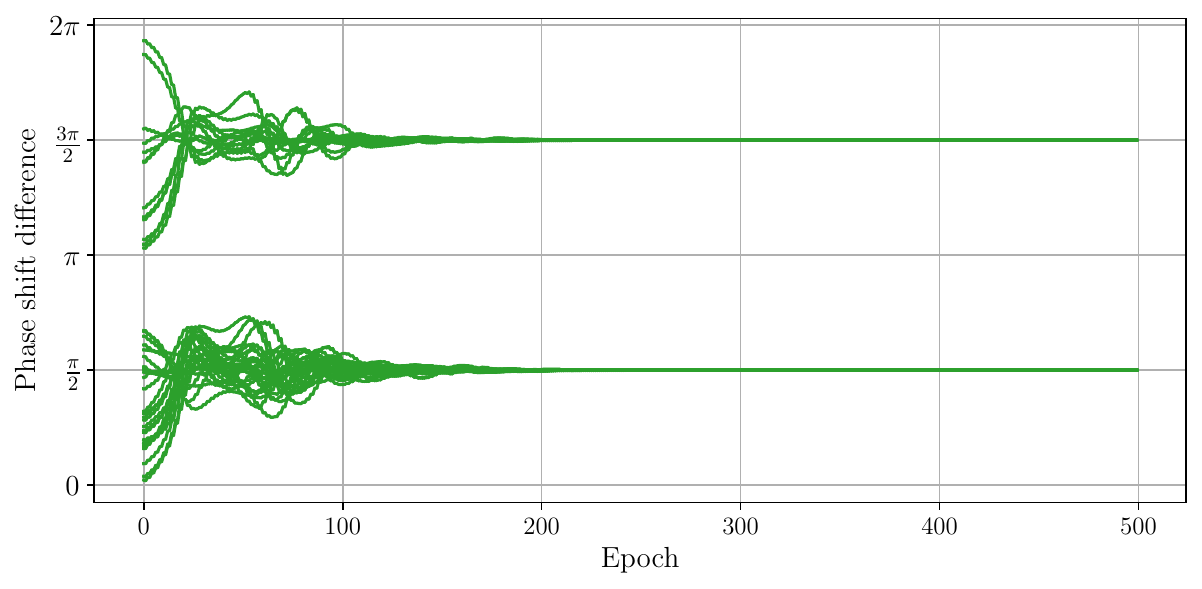}
        \caption{$N=40$.}
        \label{fig:tra1}
    \end{subfigure}
    \hfill
    \begin{subfigure}[htbp]{0.45\textwidth}
        \centering
        \includegraphics[width=\textwidth]{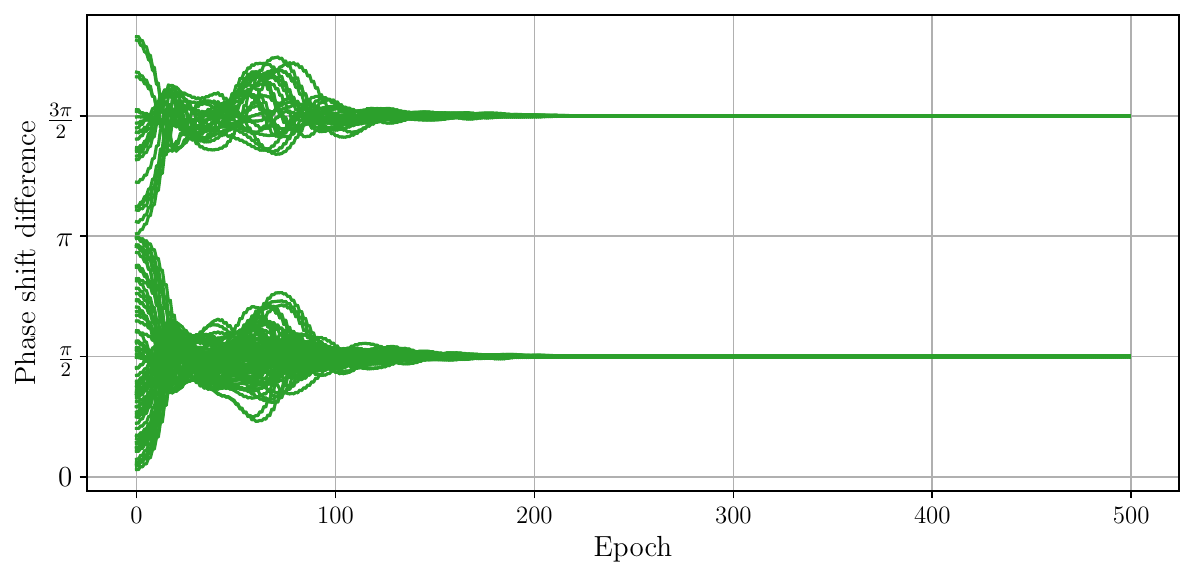}
        \caption{$N=100$.}
        \label{fig:tra2}
    \end{subfigure}
    \begin{subfigure}[htbp]{0.45\textwidth}
        \centering
        \includegraphics[width=\textwidth]{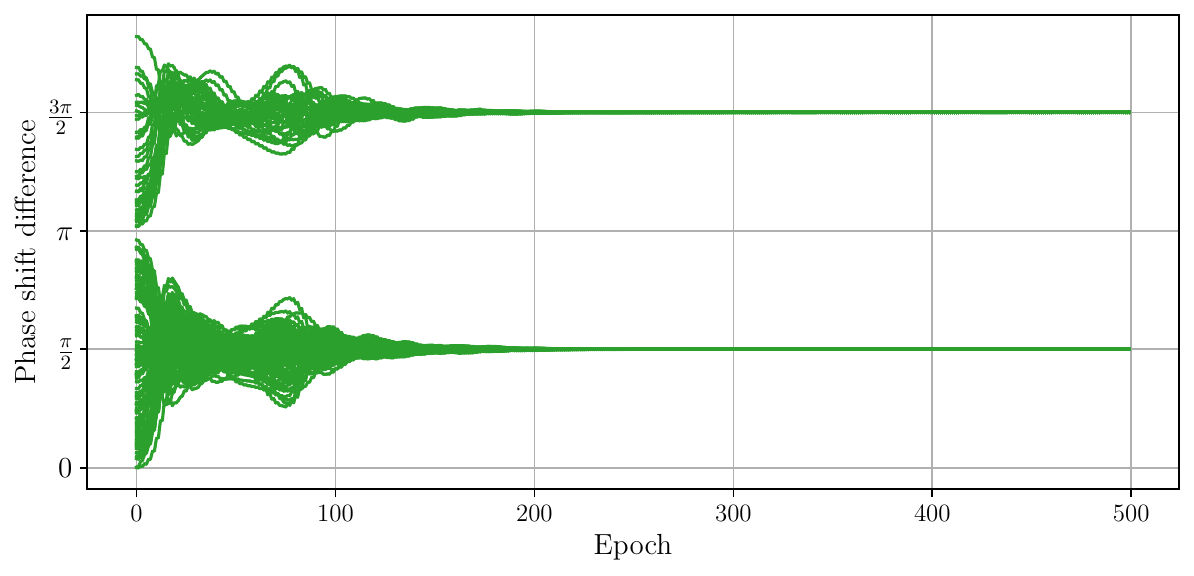}
        \caption{$N=160$.}
        \label{fig:tra3}
    \end{subfigure}
    \caption{Phase shift differences under different numbers of STAR-RIS elements.}
    \label{fig:phase_difference}
\end{figure}
In Fig. 5, we analyze the WSR performance of the proposed GML algorithm under different STAR-RIS elements. As $N$ increases, the WSR of all schemes grows monotonically, since more STAR-RIS elements provide additional spatial degrees of freedom for constructive beamforming. Moreover,  the proposed GML method for both independent-phase and coupled-phase models consistently outperform the random phase and conventional RIS baselines, which highlights the benefit of jointly optimizing the BS precoding and the STAR-RIS coefficients. Moreover, the curves of the proposed GML schemes almost overlap with those of the AO-based algorithms, indicating that the proposed approach can achieve near-optimal sum-rate performance. 


Fig. 6 illustrates the WSR performance versus the maximum transmit power $P_{\max}$ for different schemes under both independent-phase and coupled-phase models. The proposed GML-based designs for both models consistently outperform the random phase and conventional RIS baselines over the entire maximum transmit power range. Moreover, the curves of the proposed GML schemes closely follow those of the AO-based algorithms for both phase models, indicating that GML can achieve nearly the same WSR performance with much lower computational complexity. The zoomed-in inset around $P_{\max}=5$ dBm further shows that the WSR gap between GML and AO is negligible, and that the coupled-phase architecture attains a performance very close to its independent counterpart.

  
The combined effect of the numbers of BS antennas and STAR-RIS elements on the performance is depicted in Fig.~7. Both GML and AO achieve higher WSR with increasing $M$ and $N$, as more spatial resources become available for beamforming and interference mitigation. Across all configurations, the performance of GML remains nearly identical to that of AO, which verifies its scalability and effectiveness in large system settings.  


Fig. 8 provides a detailed comparison of the average CPU time of the proposed GML framework and the AO-based benchmarks under different numbers of BS antennas and STAR-RIS elements for both STAR-RIS models. For all considered settings, the runtime of every scheme increases with $M$ and $N$, since larger antenna arrays and more STAR-RIS elements enlarge the optimization dimension. However, the growth rate of the proposed GML algorithm is significantly more moderate, whose curves exhibit an almost linear scaling with $M$ for each fixed $N$, and the CPU time remains within a few tens of seconds even in the large-scale regime. In contrast, the AO-based methods show a much steeper increase in CPU time and quickly become computationally prohibitive when $M$ and $N$ grow. This behavior can be attributed to that AO-based methods rely on nested iterative updates with repeated large-dimensional matrix inversions and convex-optimization steps, leading to approximately cubic complexity, whereas the proposed GML method mainly involves lightweight forward–backward passes through neural networks whose complexity scales nearly linearly with $M$ and $N$. Moreover, the independent and coupled phase shift implementations of each algorithm exhibit very similar runtimes, indicating that the imposed phase-coupled constraint does not introduce a noticeable additional computational burden. Overall, Fig. 8 clearly demonstrates that the proposed GML framework is much more scalable and better suited for real-time implementation in large-scale STAR-RIS-assisted systems. 

In Fig. 9, we investigate the evolution of the phase shift differences between the reflection and transmission phase shifts of all STAR-RIS elements over the training epochs under the coupled phase-shift model, for $N=40$, $N=100$, and $N=160$, respectively. At the beginning of training, the phase shift differences of different elements are widely spread since the penalty term is relatively small compared to WSR. As the learning process proceeds and the penalty factor $\rho$ increases, all trajectories rapidly converge to $\pi/2$ and $3\pi/2$, which are exactly the desired phase shift differences imposed by the coupled-phase constraint. This behavior confirms that the proposed penalty-based design in the TN can effectively drive the reflection and transmission phase shifts of each STAR-RIS element to satisfy the coupling relationship, and that this property is preserved when the number of STAR-RIS elements $N$ increases from 40 to 160.

    \section{conclusion}
	In this paper, we have studied an efficient beamforming optimization for STAR-RIS-assisted MU-MISO systems under both independent-phase and coupled-phase models. First, we have formulated optimization problems for both models. Then, on this basis, we have developed a GML framework in which lightweight neural networks take gradients of the WSR with respect to the optimization variables as inputs and output refined beamforming updates, thereby eliminating the need for expensive pre-training and enabling fast adaptation. For the coupled-phase model, a penalty-based loss has been designed to gradually enforce the hardware-induced coupled-phase constraint while still prioritizing performance in the early training stages. Complexity analysis and simulations have shown that the proposed GML algorithm achieves near-AO benchmark performance with substantially reduced computational cost and runtime, and scales favorably with the number of antennas and STAR-RIS elements. Overall, the results indicated that GML is a practical and scalable solution for real-time beamforming optimization in large-scale STAR-RIS-assisted wireless networks.
    \balance
	\bibliographystyle{IEEEtran}
	\bibliography{references.bib}  

@ARTICLE{9570143,
  author={Mu, Xidong and Liu, Yuanwei and Guo, Li and Lin, Jiaru and Schober, Robert},
  journal={IEEE Trans. Wireless Commun.}, 
  title={Simultaneously Transmitting and Reflecting {(STAR)} RIS Aided Wireless Communications}, 
  year={2022},
  month={May},
  volume={21},
  number={5},
  pages={3083-3098},
  doi={10.1109/TWC.2021.3118225}
}

@ARTICLE{10623434,
  author={Zhu, Fenghao and Wang, Xinquan and Huang, Chongwen and Yang, Zhaohui and Chen, Xiaoming and Al Hammadi, Ahmed and Zhang, Zhaoyang and Yuen, Chau and Debbah, Mérouane},
  journal={IEEE Trans. Wireless Commun.}, 
  title={Robust Beamforming for {RIS}-Aided Communications: Gradient-Based Manifold Meta Learning}, 
  year={2024},
  month={Nov.},
  volume={23},
  number={11},
  pages={15945-15956},
  doi={10.1109/TWC.2024.3435023}}

@ARTICLE{8982186,
  author={Guo, Huayan and Liang, Ying-Chang and Chen, Jie and Larsson, Erik G.},
  journal={IEEE Trans. Wireless Commun.}, 
  title={Weighted Sum-Rate Maximization for Reconfigurable Intelligent Surface Aided Wireless Networks}, 
  year={2020},
  month={May},
  volume={19},
  number={5},
  pages={3064-3076}
  }

@ARTICLE{9935266,
  author={Wang, Zhaolin and Mu, Xidong and Liu, Yuanwei and Schober, Robert},
  journal={IEEE Wireless Commun. Lett.}, 
  title={Coupled Phase-Shift {STAR-RISs}: A General Optimization Framework}, 
  year={2023},
  month={Feb.},
  volume={12},
  number={2},
  pages={207-211},
  doi={10.1109/LWC.2022.3219020}}

@ARTICLE{9837935,
  author={Zhong, Ruikang and Liu, Yuanwei and Mu, Xidong and Chen, Yue and Wang, Xianbin and Hanzo, Lajos},
  journal={IEEE J. Sel. Areas Commun.}, 
  title={Hybrid Reinforcement Learning for {STAR-RISs}: A Coupled Phase-Shift Model Based Beamformer}, 
  year={2022},
  month={Sep.},
  volume={40},
  number={9},
  pages={2556-2569},
  doi={10.1109/JSAC.2022.3192053}}

@ARTICLE{8741198,
  author={Huang, Chongwen and Zappone, Alessio and Alexandropoulos, George C. and Debbah, Mérouane and Yuen, Chau},
  journal={IEEE Trans. Wireless Commun.}, 
  title={Reconfigurable Intelligent Surfaces for Energy Efficiency in Wireless Communication}, 
  year={2019},
  month={Aug.},
  volume={18},
  number={8},
  pages={4157-4170},
  doi={10.1109/TWC.2019.2922609}}

@ARTICLE{9424177,
  author={Liu, Yuanwei and Liu, Xiao and Mu, Xidong and Hou, Tianwei and Xu, Jiaqi and Di Renzo, Marco and Al-Dhahir, Naofal},
  journal={IEEE Commun. Surveys Tuts.}, 
  title={Reconfigurable Intelligent Surfaces: Principles and Opportunities}, 
  year={2021},
  month={thirdquarter},
  volume={23},
  number={3},
  pages={1546-1577},
  doi={10.1109/COMST.2021.3077737}}

@ARTICLE{9437234,
  author={Xu, Jiaqi and Liu, Yuanwei and Mu, Xidong and Dobre, Octavia A.},
  journal={IEEE Commun. Lett}, 
  title={{STAR-RISs}: Simultaneous Transmitting and Reflecting Reconfigurable Intelligent Surfaces}, 
  year={2021},
  month={Sep.},
  volume={25},
  number={9},
  pages={3134-3138},
  doi={10.1109/LCOMM.2021.3082214}}

@ARTICLE{9690478,
  author={Liu, Yuanwei and Mu, Xidong and Xu, Jiaqi and Schober, Robert and Hao, Yang and Poor, H. Vincent and Hanzo, Lajos},
  journal={IEEE Wireless Commun.}, 
  title={{STAR}: Simultaneous Transmission and Reflection for {360°} Coverage by Intelligent Surfaces}, 
  year={2021},
  month={Dec.},
  volume={28},
  number={6},
  pages={102-109},
  doi={10.1109/MWC.001.2100191}}

@ARTICLE{9629335,
  author={Niu, Hehao and Chu, Zheng and Zhou, Fuhui and Xiao, Pei and Al-Dhahir, Naofal},
  journal={IEEE Trans. Veh. Technol.}, 
  title={Weighted Sum Rate Optimization for {STAR-RIS}-Assisted {MIMO} System}, 
  year={2022},
  month={Feb.},
  volume={71},
  number={2},
  pages={2122-2127},
  doi={10.1109/TVT.2021.3131568}}

@ARTICLE{10316600,
  author={Song, Yaxin and Xu, Shaoyi and Xu, Rongtao and Ai, Bo},
  journal={IEEE Trans. Veh. Technol.}, 
  title={Weighted Sum-Rate Maximization for Multi-{STAR-RIS}-Assisted {mmWave} Cell-Free Networks}, 
  year={2024},
  month={Apr.},
  volume={73},
  number={4},
  pages={5304-5320},
  doi={10.1109/TVT.2023.3332334}}

@ARTICLE{10475146,
  author={Luo, Xiaomei and Jiang, Zhuochen and Xu, Fan and Li, Xiaoyang and Zhu, Guangxu and Shen, Kaiming},
  journal={IEEE Wireless Commun. Lett.}, 
  title={Sum-Rate Maximization for {STAR-RIS}-Assisted Multi-User Networks With Hardware Impairments}, 
  year={2024},
  month={May},
  volume={13},
  number={5},
  pages={1503-1507},
  doi={10.1109/LWC.2024.3378846}}

@ARTICLE{10188900,
  author={Wang, Chao and Wang, Cheng-Cai and Li, Zan and Ng, Derrick Wing Kwan and Wong, Kai-Kit and Al-Dhahir, Naofal and Niyato, Dusit},
  journal={IEEE Trans. Inf. Forensics Secur.}, 
  title={{STAR-RIS}-Enabled Secure Dual-Functional Radar-Communications: Joint Waveform and Reflective Beamforming Optimization}, 
  year={2023},
  month={Jul.},
  volume={18},
  number={},
  pages={4577-4592},
  doi={10.1109/TIFS.2023.3297452}}

@ARTICLE{10620247,
  author={Wang, Xinquan and Zhu, Fenghao and Huang, Chongwen and Alhammadi, Ahmed and Bader, Faouzi and Zhang, Zhaoyang and Yuen, Chau and Debbah, Mérouane},
  journal={IEEE Wireless Commun. Lett.}, 
  title={Robust Beamforming With Gradient-Based Liquid Neural Network}, 
  year={2024},
  month={Nov.},
  volume={13},
  number={11},
  pages={3020-3024},
  doi={10.1109/LWC.2024.3436576}}

@INPROCEEDINGS{10901346,
  author={Gao, Zhichao and Zhong, Ruikang and Mu, Xidong and Liu, Ju and Liu, Yuanwei},
  booktitle={IEEE Global Commun. Conf.}, 
  title={Beamforming Based on {DRL} for {STAR-RIS} Aided Cell-Free Massive {MIMO} Network}, 
  year={2024},
  month={Dec.},
  address={Cape Town, South Africa},
  volume={},
  number={},
  pages={1173-1178},
  doi={10.1109/GLOBECOM52923.2024.10901346}}

@ARTICLE{10565781,
  author={Zhang, Jifa and Gong, Shiqi and Lu, Weidang and Xing, Chengwen and Zhao, Nan and Ng, Derrick Wing Kwan and Niyato, Dusit},
  journal={IEEE Trans. Wireless Commun.}, 
  title={Joint Design for {STAR-RIS} Aided {ISAC}: Decoupling or Learning}, 
  year={2024},
  month={Oct.},
  volume={23},
  number={10},
  pages={14365-14379},
  doi={10.1109/TWC.2024.3413089}}

@ARTICLE{10021676,
  author={Li, Kaiyue and Huang, Chong and Gong, Yu and Chen, Gaojie},
  journal={IEEE Wireless Commun. Lett.}, 
  title={Double Deep Learning for Joint Phase-Shift and Beamforming Based on Cascaded Channels in {RIS}-Assisted {MIMO} Networks}, 
  year={2023},
  volume={12},
  number={4},
  pages={659-663},
  doi={10.1109/LWC.2023.3238073}}

@ARTICLE{10299716,
  author={Yuan, Qijiang and Xiao, Lixia and He, Chunlin and Xiao, Pei and Jiang, Tao},
  journal={IEEE Wireless Commun. Lett.}, 
  title={Deep Learning-Based Hybrid Precoding for {RIS}-Aided Broadband Terahertz Communication Systems in the Face of Beam Squint}, 
  year={2024},
  volume={13},
  number={2},
  pages={303-307},
  doi={10.1109/LWC.2023.3327769}}

@INPROCEEDINGS{11148930,
  author={Parihar, Abhinav Singh and Kumar, Anand and Singh, Keshav and Bhatia, Vimal},
  booktitle={Proc. IEEE/CIC International Conference on Communications in China (ICCC)}, 
  title={Deep Learning for Blocklength Optimization in Fully Connected {RIS}-Aided Short-Packet {NOMA}}, 
  year={2025},
  address={Shanghai, China},
  volume={},
  number={},
  pages={1-6},
  doi={10.1109/ICCC65529.2025.11148930}}

@INPROCEEDINGS{10622978,
  author={Wang, Xinquan and Zhu, Fenghao and Zhou, Qianyun and Yu, Qihao and Huang, Chongwen and Alhammadi, Ahmed and Zhang, Zhaoyang and Yuen, Chau and Debbah, Mérouane},
  booktitle={IEEE International Conference on Communications}, 
  title={Energy-Efficient Beamforming for {RISs}-Aided Communications: Gradient Based Meta Learning}, 
  year={2024},
  address={Denver, USA},
  volume={},
  number={},
  pages={3464-3469},
  doi={10.1109/ICC51166.2024.10622978}}
\end{document}